\documentclass[review]{elsarticle}
\usepackage{amsmath}
\usepackage{amsfonts}
\usepackage{longtable}
\usepackage{graphicx}

\newcommand{\mh}{\mathcal{H}}
\newcommand{\beq}{\begin{equation}}
\newcommand{\eeq}{\end{equation}}
\def\sinc#1{\mbox{ sinc}(#1 \Delta \theta)}

\begin{document}

\begin{frontmatter}

\title{Analytical results on the magnetization of the Hamiltonian Mean Field model}

\author[SOL]{R. Bachelard}
\author[CPT]{C. Chandre}
\author[FLO]{A. Ciani}
\author[FLO]{D. Fanelli}
\author[TOK]{Y. Y. Yamaguchi}
\address[SOL]{Synchrotron Soleil, L'Orme des Merisiers, Saint-Aubin - BP 48, F-91192 Gif-sur-Yvette cedex, France}
\address[CPT]{Centre de Physique Th\'eorique, CNRS - Aix-Marseille Universit\'es, Campus de Luminy, case 907, F-13288 Marseille cedex 09, France}
\address[FLO]{Dipartimento di Energetica \char`\"{}Sergio Stecco\char`\"{}, 
Universit\'a di Firenze, via s. Marta 3, 50139 Firenze, Italia and Centro interdipartimentale
per lo Studio delle Dinamiche Complesse (CSDC) and INFN}
\address[TOK]{Department of Applied Mathematics and Physics, Graduate School of Informatics, Kyoto University, 606-8501, Kyoto, Japan}

\begin{abstract}
The violent relaxation and the metastable states of the Hamiltonian Mean-Field model, a paradigmatic system of long-range interactions, is studied using a Hamiltonian 
formalism. Rigorous results are derived algebraically for the time evolution of selected macroscopic observables, e.g., the global magnetization. The  
high and low energy limits are investigated and the analytical predictions are compared with direct $N$-body simulations. The method we use enables us to 
re-interpret the out-of-equilibrium phase transition separating magnetized and (almost) unmagnetized regimes. 
\end{abstract}


\end{frontmatter}

\section{Introduction}

Systems with long-range interactions~\cite{dauxois,campa08} exhibit a fascinating feature of metastability: Starting from out-of-equilibrium initial conditions, 
the system violently relaxes toward a metastable state, often called Quasi-Stationary State (QSS). In this regime, macroscopic quantities reach values 
which substantially differ from the corresponding thermodynamic equilibrium configuration. Although the QSS are only transient regimes, their lifetime have 
been shown to diverge with the number of bodies in interaction~\cite{antoni}. For this reason they possibly correspond to the solely accessible
experimental regimes. 

We consider a paradigmatic system with long-range interactions, the Hamiltonian Mean-Field (HMF)~\cite{antoni} where particles on a circle are 
collectively interacting through a cosine-like mean-field potential. After a fast relaxation, the system typically enters a metastable regime in which the particles
either aggregate into a large cluster (magnetized phase), or they spread almost homogeneously around the circle (unmagnetized or homogeneous phase). In particular, an 
out-of-equilibrium phase transition 
between these two states occurs when the parameters of the initial conditions are varied~\cite{antoniazzi07}. 

In this paper, we focus on both the violent relaxation process and the subsequent QSS regime. We use an algebraic framework based on a Hamiltonian formulation of the Vlasov equation for the HMF model. This Vlasov equation rules the evolution of the single
particle distribution function in phase space (as a kinetic equation) and naturally arises when investigating the continuous version of the HMF model. As in the limit of infinite 
number of particles the system gets permanently frozen in the QSS phase, it is customarily believed that QSS can be interpreted as equilibria of the
Vlasov equation.  We exploit a Hamiltonian formalism of this Vlasov equation to derive analytical expressions for the global magnetization as function of time. This magnetization measures the aggregation of the particles on the circle. It is 
a macroscopic observable which is directly influenced by the microscopic, single particle trajectory. It is in general particularly cumbersome 
to bridge the gap between the microscopic realm of the many-body interacting constituents and the macroscopic world of collective dynamics. 

Using an expansion provided by the Hamiltonian framework, we here obtain rigorous results on the time expansion of relevant observables. These results are compared with direct numerical simulation. We consider in particular the high and low energy regimes which allow some simplifications in the expansions. In addition, we characterize the aforementioned 
out-of-equilibrium phase transition which occurs in an intermediate energy range. This is achieved by monitoring the initial relaxation of the magnetization, as a function of relevant parameters of the initial distribution. The parameter space is hence partitioned into two regions, depending on the magnetization amount, a result 
which positively correlates with direct numerics~\cite{antoniazzi07}. 

The paper is organized as follows: In Sec.~\ref{sec2} we will review the discrete HMF model, presents its continuous counterpart and discuss the basic of the bracket expansion
method. Section~\ref{sec3} is devoted to the presentation of the analytical results, with special emphasis to the high and low energy regimes. The out-of-equilibrium phase transition issue is also
addressed. Comparison with direct simulations is provided to substantiate the accuracy of our predictions.

\section{Model and methods}
\label{sec2}

\subsection{Lie-Poisson structure of the Vlasov equation}

\begin{figure}
   \center
   \includegraphics[width=8cm]{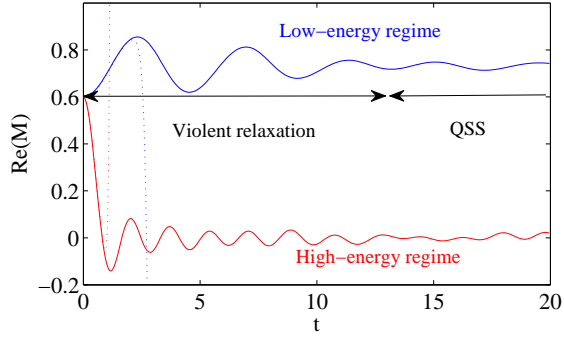}\\
   \caption{Real part of the magnetization given by Eq.~(\ref{eqn:magn}) as a function of time obtained by integrating the dynamics 
   given by Eq.~(\ref{eq:evol}) for $M_0=0.6$. The system reaches either a finite-magnetization for low energies ($U=0.4$, in blue), 
   or a low-magnetization for high energies ($U=3$, in red). The plain lines refer to $N$-body simulations (with $N=10000$), while 
   the dotted lines come from the predictions given by Eq.~(\ref{eq:taylor}) for $k_0=20$.}
   \label{fig:2regimes}
\end{figure}

\begin{figure}
\center
\begin{tabular}{c}
\includegraphics[width=8cm]{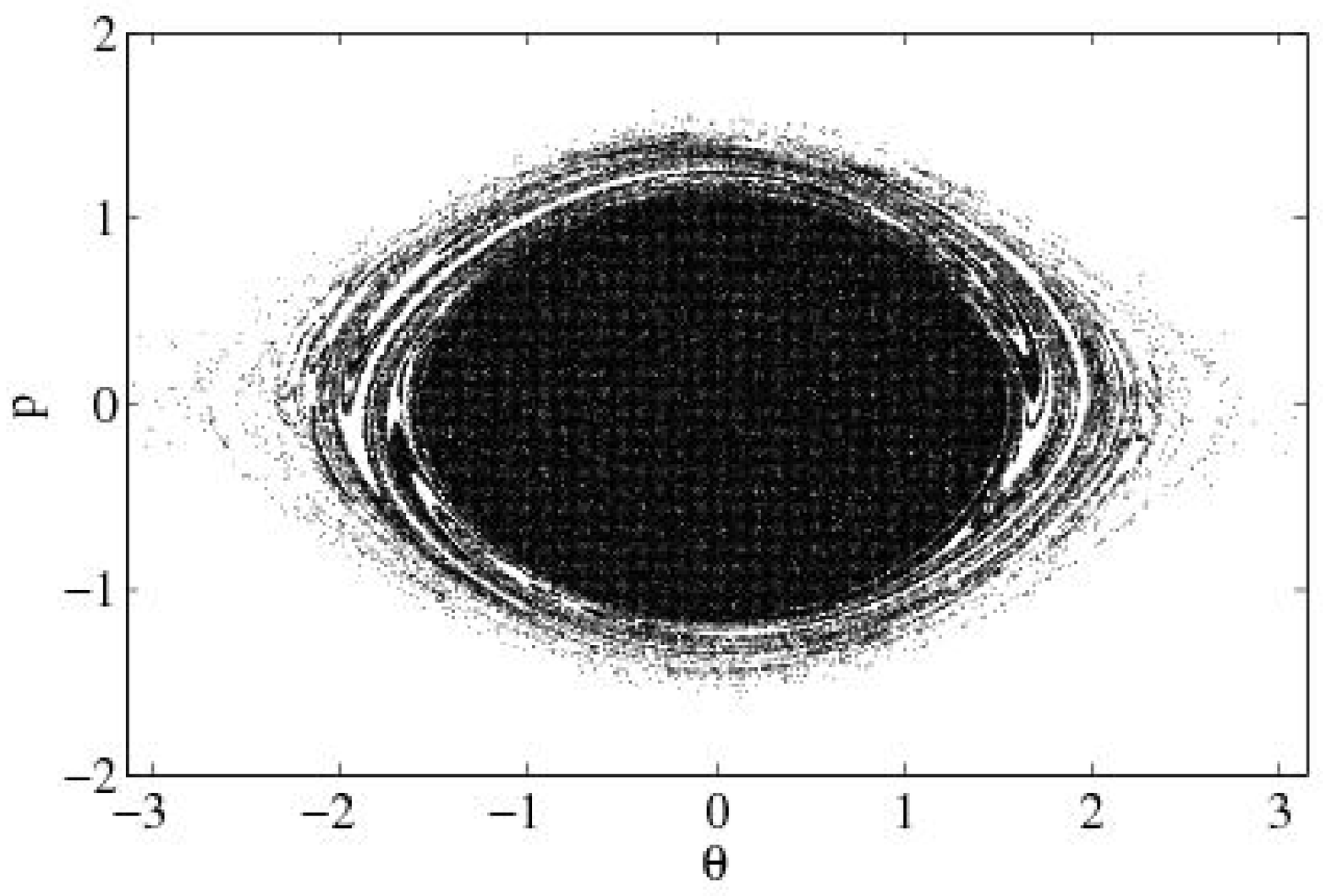}
\\ \includegraphics[width=8cm]{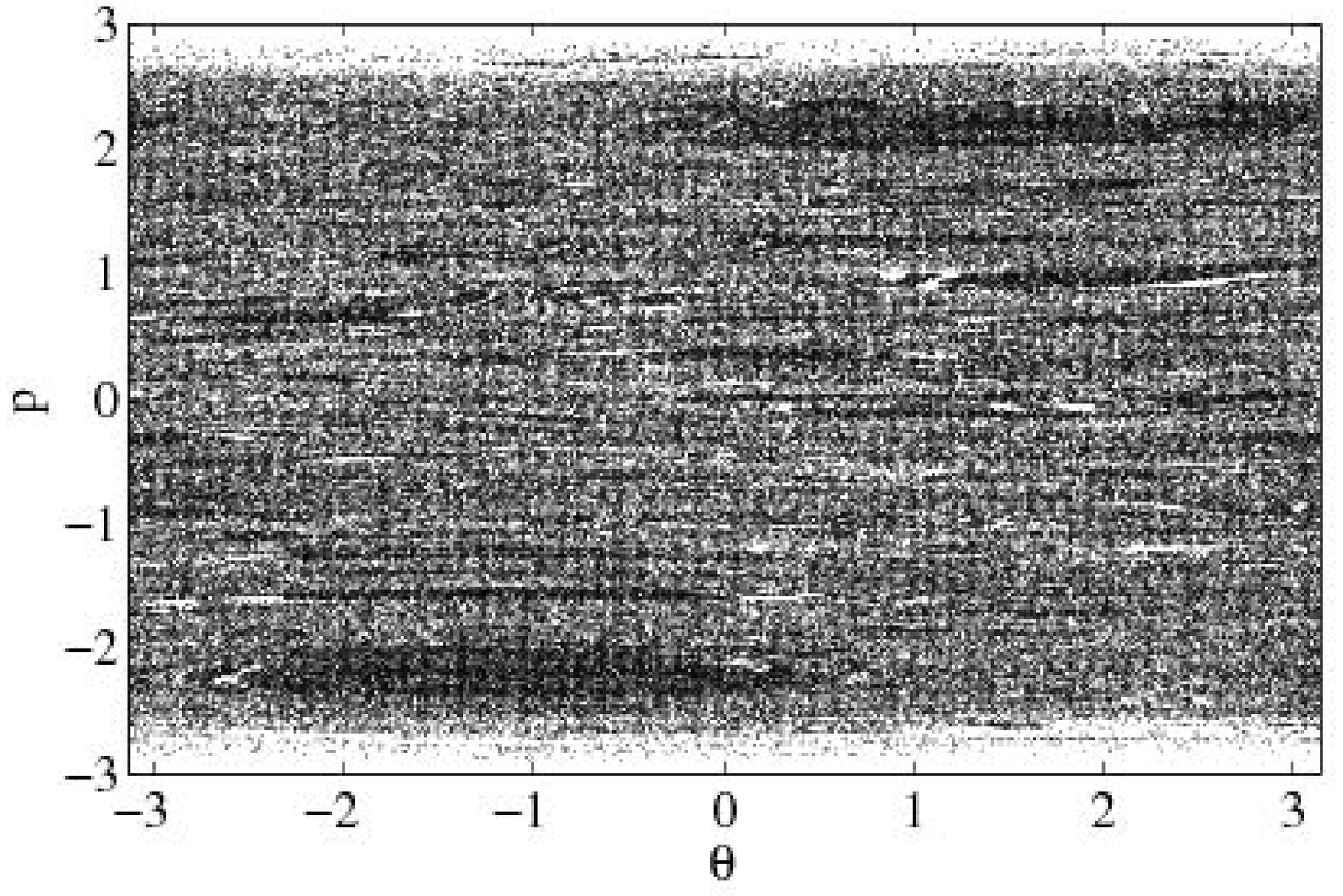}
\end{tabular}
\caption{Phase space portrait of the system~(\ref{eqn:HamN}) once saturation has been reached for $M_0=0.6$: The low energy regime ($U=0.5$, upper panel) 
is characterized by one large cluster of particles, whereas for higher energies ($U=1.7$, lower panel), phase space is quite 
homogeneous, except for two clusters $at \pm 2.2$ (i.e. moving in opposite directions).}
\label{fig:phasespace}
\end{figure}

We consider $N$ particles interacting on a circle with the following Hamiltonian:
\begin{equation}
\label{eqn:HamN}
H=\sum_{i=1}^N \left[ \frac{p_i^2}{2}+\frac{1}{2N}\sum_{j=1}^N[1-\cos(\theta_i-\theta_j)]\right],
\end{equation}
where $(\theta_i,p_i)$ are canonically conjugate variables which means that the Poisson bracket giving the dynamics (Hamilton's equations) is given by
$$
\{F,G\}=\sum_{i=1}^N\left(\frac{\partial F}{\partial p_i}\frac{\partial G}{\partial \theta_i} -
\frac{\partial F}{\partial \theta_i}\frac{\partial G}{\partial p_i}\right).
$$
In the continuous limit, we consider an Eulerian description of the system which gives the dynamical evolution of the distribution of particles $f(\theta,p;t)$ 
in phase space via the following Vlasov equation:
\begin{equation}
\label{eqn:Vlasov}
\frac{\partial f}{\partial t}=-p\frac{\partial f}{\partial \theta}+\frac{dV[f]}{d\theta}\frac{\partial f}{\partial p},
\end{equation}
where $V[f](\theta)=1-M_x[f]\cos\theta -M_y[f]\sin\theta$. The magnetization $M[f]=M_x+i M_y$ is defined as 
\begin{equation}
\label{eqn:magn}
M[f]=\iint f e^{i\theta}d\theta dp,
\end{equation}
where the integrals are taken over $[-\pi, \pi]\times {\mathbb R}$.
Equation~(\ref{eqn:Vlasov}) can be cast into a Hamiltonian form where the (infinite dimensional) phase space is composed of the functions $f(\theta,p)$ of $]-\pi,\pi]\times{\mathbb R}$. The Hamiltonian is given by  
\begin{equation}
H[f]=\iint f\frac{p^{2}}{2}d\theta dp-\frac{M_x[f]^2+M_y[f]^2-1}{2}, \label{eq:ham}
\end{equation}
and the associated Lie-Poisson bracket by
\begin{equation}
\{F,G\}=\iint f \bigg( \frac{\partial}{\partial p}\frac{\delta F}{\delta f}  \frac{\partial}{\partial \theta} \frac{\delta G}{\delta f} 
-\frac{\partial}{\partial \theta}\frac{\delta F}{\delta f}  \frac{\partial}{\partial p} \frac{\delta G}{\delta f} \bigg)  d\theta dp, \label{eq:bra}\end{equation}
for $F$ and $G$ two observables (that is, functionals of $f$). The functional derivatives $\delta F/\delta f$ are computed following the expansion~:
$$
F[f+\varphi]-F[f]=\iint  \frac{\delta F}{\delta f} \varphi d\theta dp +O(\varphi^2).
$$
The Poisson bracket~(\ref{eq:bra}) satisfies several properties: bilinearity, Leibniz rule and Jacobi identity (for more details, see Refs.~\cite{abraB78,morr82}). Its Casimir invariants are given by
$$
C[f]=\iint c(f) d\theta dp ,
$$
where $c(f)$ is any function of $f(\theta,p)$. In particular, the total distribution $\iint f d\theta dp$ is one of such Casimir invariants and hence is conserved by the flow. 
The evolution of any observable $F[f]$ is then given by
\begin{equation}
\dot{F}=\{H,F\}. \label{eq:evol}
\end{equation}
For instance, for $F[f]=f(\theta,p)$, we recover Eq.~(\ref{eqn:Vlasov}). Another convenient observable to study is the magnetization $M[f]$ given by Eq.~(\ref{eqn:magn}): It quantifies the spatial aggregation of the particles. At low energies, the magnetization typically relaxes until it reaches an out-of-equilibrium plateau, around which it fluctuates (see Fig.~\ref{fig:2regimes}). In this case, the particles are trapped into the large resonance created by the finite magnetization, hence the name ``magnetized state'' (see upper panel of Fig.~\ref{fig:phasespace}). At high energies, the magnetization falls and fluctuates around zero (see Fig.~\ref{fig:2regimes}), which means that the particles failed to organize collectively. This is called the ``homogeneous phase'' (see lower panel of Fig.~\ref{fig:phasespace}).

The dynamics given by Eq.~(\ref{eq:evol}) is deduced from the linear operator ${\mathcal H}$. From the evaluation of the functional derivative of $H$ with respect to $f$
$$
\frac{\delta H}{\delta f}=\frac{p^2}{2}-M_x[f]\cos\theta-M_y[f]\sin\theta,
$$
we get the expression of $\mh$~:
\begin{eqnarray}
\mh &\equiv& \{H,.\} \nonumber
\\ &=& \iint d\theta dp f \left(p\frac{\partial}{\partial\theta}  +\frac{Me^{-i\theta}-M^*e^{i\theta}}{2i}\frac{\partial}{\partial p}  \right)\frac{\delta}{\delta f} . \label{eq:mh}
\end{eqnarray}

In the algebraic computations that follows, we make an explicit use of the linearity of $\mh$ and Leibniz rule:
\begin{eqnarray*}
&& \mh (F+\alpha G)=\mh F +\alpha \mh G,\\
&& \mh (FG)=F\mh G+(\mh F) G.
\end{eqnarray*}

{\em $N$-body simulations (Lagrangian point of view)}: In order to compare the algebraic results with numerical ones, we integrate 
Eq.~(\ref{eq:evol}) via $N$-body simulations, which are obtained by considering a Klimontovitch \cite{nicholson} distribution of particles
$$
f(\theta,p;t)=\frac{1}{N}\sum_{i=1}^N \delta\left(\theta -\theta_i(t) \right)\delta\left(p -p_i(t) \right),
$$
whose dynamics is ultimately reduced to Hamiltonian~(\ref{eqn:HamN}).
Such simulations are used with a large number of particles (typically $N=10^5$) such that the intermediate regime experienced by the 
$N$-body simulations is close to the behavior of the Vlasov equations (at least for some observables like the magnetization)~\cite{anto07}.

\subsection{Bracket method}

The evolution of a selected observable $F[f]$ given by Eq.~(\ref{eq:evol}) is obtained formally from the operator ${\mathcal H}$ as
\beq F[f](t)=e^{t\mh}F [f_0]\equiv\sum_{k=0}^{\infty}\frac{t^k}{k!}\mh^{k}F [f_0], \eeq
where $f_0$ is the initial distribution.
Here we compute a finite number of terms in this series in order to obtain a Taylor expansion for the solution of the dynamics of $F$:
\begin{equation}
\label{eq:taylor}
F[f](t)\approx \sum_{k=0}^{k_0} \frac{t^k}{k!}\mh^kF [f_0],
\end{equation}
where $k_0$ is the truncation parameter. Of course, this approximation is accurate up to some time $t$ depending on $k_0$. 
A convergence over longer times is expected for increasing $k_0$. Furthermore, the coefficients of the series $\mh^k F$ are obtained recursively 
by applying $\mh$ on the previous term $\mh^{k-1}F$. Finally, note that Eq.~(\ref{eq:taylor}) yields an explicit dependence on the initial conditions, 
and it is not restricted to close-to-equilibrium initial conditions, thus being a useful tool to investigate the far-from-equilibrium violent 
relaxation of the system.

We consider the subspace of functions composed by sums and products of the following elements (which are also functionals of $f$):
$$
b_{n,m}[f]=\iint d\theta dp f e^{in\theta} p^{m},
$$ 
where $(n,m)\in{\mathbb Z}\times{\mathbb N}$. We notice that the main observables of the system such as the $n$-th order magnetization $M_n=\iint e^{in\theta}f d\theta dp$ or the momenta $P_m=\iint p^m f d\theta dp$ of the system belong to this family. Furthermore this family is stable by the action of $\mh$ given by 
\beq 
\label{eqn:recur}
\mh b_{n,m}=inb_{n,m+1}+\frac{m}{2i}(b_{1,0}b_{n-1,m-1}-b_{-1,0}b_{n+1,m-1}). \eeq
We notice that only positive values of $m$ are involved in the iterations of the recursion relation since the second term is proportional to $m$.   
Taking into account the linearity and the Leibniz rule for $\mh$ mentioned in the previous section, the derivation of the short-time evolution~(\ref{eq:taylor}) 
of a given observable $F$ is computed algebraically as a sum of products of elements $b_{n,m}$. For instance, the magnetization is given by $M=b_{1,0}$ and its 
first order evolution is obtained from Eq.~(\ref{eqn:recur}):
\begin{eqnarray*}
M(t)&=&b_{1,0}[f_0]+t \mh b_{1,0}[f_0]+\frac{t^2}{2}\mh^2 b_{1,0}[f_0]+O(t^3),\\
&=&b_{1,0}[f_0]+it b_{1,1}[f_0]+i\frac{t^2}{2}\mh b_{1,1}[f_0]+O(t^3),\\
&=& b_{1,0}[f_0]+it b_{1,1}[f_0]\\
&&  +\frac{t^2}{2}\left( -b_{1,2}[f_0]+\frac{1}{2}(b_{1,0}[f_0]-b_{-1,0}[f_0]b_{2,0}[f_0])\right)\\
&& +O(t^3).
\end{eqnarray*}
Of course, a satisfying approximation of the time evolution of any observable needs a 
large number of terms in the expansion~(\ref{eq:taylor}). At a given time $t$, the number of terms necessary to obtain a reasonably good approximation 
of the dynamics depends on the initial distribution $f_0$ as it is shown in Fig.~\ref{fig:2regimes} where, at low energies, the accuracy extends to longer 
times than at high energies. In addition, we need to specify the initial distribution which will be used to compute $b_{n,m}(0)$ 
necessary to complete the computation of the approximate evolution. In the following sections, we use a waterbag distribution as initial condition. 

\subsection{Initial conditions}

The waterbag initial distribution is a uniform distribution over a rectangle in phase space corresponding to the points $(\theta,p)\in[-\Delta\theta, 
\Delta \theta]\times [-\Delta p, \Delta p]$. The distribution $f_0(\theta,p)$ is equal to $1/(4\Delta \theta \Delta p)$ if $(\theta,p)\in[-\Delta\theta, 
\Delta \theta]\times [-\Delta p, \Delta p]$ and zero otherwise. The values of $b_{n,m}$ at $t=0$ can be computed explicitly in this case and are equal to
$$
b_{n,m}(0)=\frac{(\Delta p)^{m+1}-(-\Delta p)^{m+1}}{2(m+1)\Delta p}{\rm sinc} (n\Delta \theta),
$$
where ${\rm sinc}(\cdot) = \sin(\cdot)/(\cdot)$. 
In particular, we notice that $b_{n,m}(0)=0$ for $m$ odd. 
The waterbag is characterized by two parameters $(\Delta \theta,\Delta p)$. Instead we consider the initial magnetization $M_0$ and the energy $U$ to label the initial conditions:
\begin{eqnarray*}
&& M_0\equiv b_{1,0}(0)={\rm sinc} (\Delta \theta),\\
&& U\equiv \frac{1}{2}\left(b_{0,2}-b_{1,0} b_{-1,0}+1\right)=\frac{\Delta p ^2}{6}-\frac{M_0^2-1}{2}.
\end{eqnarray*}
In the following, we investigate the high energy $U\gg 1$ and low energy $U\ll 1$ limits for the initial distribution. 

\section{Analytical results}
\label{sec3}

\begin{figure}
\center
\includegraphics[width=8cm]{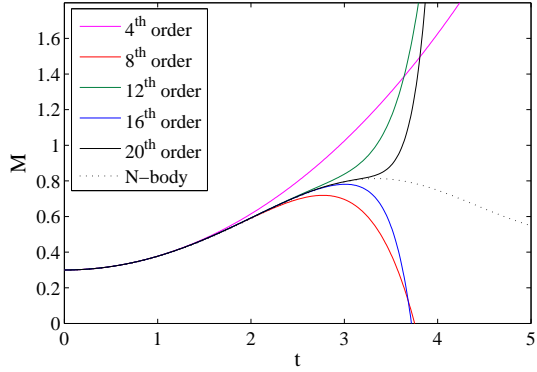}
\caption{Magnetization $M(t)$ versus time for $M_0=0.2$ and $U=0.6$ obtained from $N$-body simulations (dotted black curve) and using the algebraic expansions at various orders from the 4th to the 20th order. }
\label{fig:magnDO}
\end{figure}

The first terms of the expansion  for the magnetization $M(t)$ given by Eq.~(\ref{eq:taylor}) (for a waterbag initial distribution) are listed in Tab.~\ref{tab:1} up to 
sixth order in time. We notice that the number of terms in the expansion increases exponentially, making such expressions difficult to handle in 
practice. In Fig.~\ref{fig:2regimes}, we notice that even with $k_0=20$ which involves approximately one thousand terms, a good agreement is observed 
only up to $t=2$. In Fig.~\ref{fig:magnDO}, the algebraic expressions for the magnetization obtained by Eq.~(\ref{eq:taylor}) are plotted at different orders.  
Other than the initial regime, if one is interested in the intermediate regimes, the only hope is to find the governing rules behind this algebraic 
computations in order to draw some conclusions. This is the case for the low and high energy limits where the leading terms 
of the expansion can be extracted to all orders. These simplifications allow us to derive some dynamical properties of the system.  

\begin{longtable}{c|c|l}
\caption{\label{tab:1}First terms in the expansion of the magnetization $M(t)$ given by Eq.~(\ref{eq:taylor}) for the waterbag initial distribution.}\\

\hline\hline

$t^2/2!$ & $\Delta p^0$ & $(1-\sinc{2})M_0/2$ \\

 & $\Delta p^{2}$ & $-M_0/3 $ \\

 \hline

 $t^4/4!$ & $\Delta p^0$ &
 $(1-2\sinc{2}+\sinc{2}^2-4M_0^2+4M_0\sinc{3})M_0/4$\\

 & $\Delta p^2$ & $-2\left(3\sinc{2}+1\right) M_{0}/3$ \\


 & $\Delta p^4$ & $M_0/5$ \\

 \hline

 $t^6/6!$ & $\Delta p^0$ & $(1+3\sinc{2}+64 M_0^2+3\sinc{2}^2+26 M_0\sinc{3}$\\
 & & $+98 M_0^2\sinc{2}-\sinc{2}^3-34M_0^2\sinc{4}$\\
 & & $-26M_0\sinc{2}\sinc{3})M_0/8$\\

 & $\Delta p^2$ & $(-202\sinc{2} M_{0}
           -51\sinc{2}^2
           +58\sinc{2}
           +138 M_{0}^{2}$\\
  & & $-7) M_{0}/12$ \\


 & $\Delta p^4$ &
 $\left(-239\sinc{2}    +23\right) M_{0}/30$\\

 & $\Delta p^6$ & $-M_0/7$ \\

\hline\hline
\end{longtable}

\subsection{High-energy limit}
\label{sec:HE}

\begin{figure}
   \center
   \includegraphics[width=8cm]{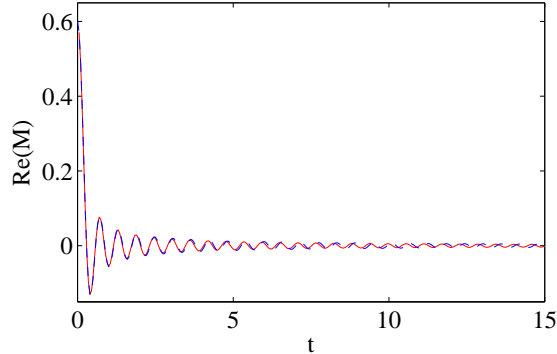}\\
   \caption{Real part of the magnetization $M(t)$ versus time for $M_0=0.6$ and $U=20$. The dashed red line refers to direct simulations, while the solid blue one stands for the approximate solution~(\ref{eq:HE}).}
   \label{fig:HE}
\end{figure}

First we consider the high energy limit, which corresponds to $\Delta p \gg 1$ in the initial waterbag. In this regime, since the kinetic term is dominant, 
the dynamics is driven by the reduced Liouville operator, which takes into account only the kinetic term
\beq \mh_{HE}=\iint d\theta dp f p\frac{\partial}{\partial \theta}\frac{\delta}{\delta f}. \eeq
From Eq.~(\ref{eqn:recur}), the successive actions of $\mh$ on $b_{n,m}$ is given by
$$
\mh^k b_{n,m}=(in)^k b_{n,m+k}.
$$
For the waterbag initial distribution, it is straightforward to deduce the evolution of the magnetization of order $n$:
\beq M_n(t)=M_n(0) {\rm sinc}{(n\Delta p t)}.\label{eq:HE}\eeq
The magnetization envelop exhibits a slow decay (as $1/(\Delta p t)$) towards the asymptotic (equilibrium)
state $M=0$ (see Fig.~\ref{fig:HE}). 
  
The profile obtained from $N$-body simulations is correctly interpolated over a finite 
time window by Eq.~(\ref{eq:HE}). As $U$ is increased, the agreement gets better, even if deviations from Eq.~(\ref{eq:HE}) are observed at later times. 
Such a discrepancy is due to the cumulative effects of the neglected contributions in $\Delta p$ (see Tab.~\ref{tab:1}). 

\begin{figure}
   \center
   \includegraphics[width=7cm]{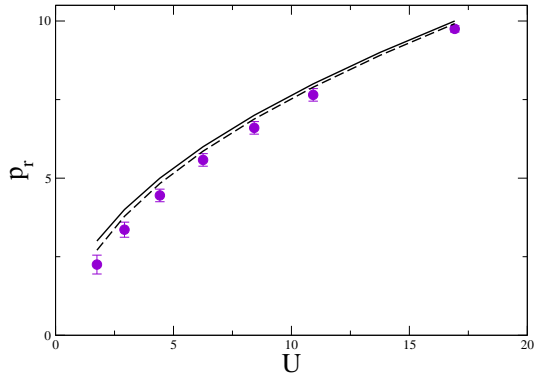}
   \caption{The velocity of the resonances (in the high energy regime) as a function of the energy $U$ for $M_0=0.7$. The circles refer to the velocities 
   obtained numerically using $N$-body simulations. The vertical bars delimit the width of the resonances in the $p$-direction. 
   The solid line is given by Eq.~(\ref{pr}), and the dashed line by Eq.~(\ref{sinc}).}
   \label{fig7}
\end{figure}

It was reported in Ref.~\cite{bach08} that for large values of the energy, and for any given 
initial magnetization, two large resonances spontaneously develop and effectively divide the available phase space into independent regions. 
Such resonances move in opposite directions, over the unit circle. Their velocity $p_r$ is identical in modulus and tends to grow as the energy is
increased. The 
magnetization $M(t)$ is mostly influenced by the instantaneous positions of the resonances. A snapshot of the positions of the particles obtained using $N$-body simulations 
is depicted in Fig.~\ref{fig:phasespace} (lower panel). It reveals the two resonances moving in opposite directions (with velocity $p_r$). The maxima of $M(t)$ are obtained when the two resonances are aligned since the bunching 
of particles is maximum in this case. During two successive maxima of $M(t)$ each cluster travels on a segment of length $2\pi$ in $\theta$, which 
takes a time $2\pi/p_r$. On the other hand, according to Eq.~(\ref{eq:HE}), two successive bumps in the magnetization are separated by a time 
interval $2\pi/{\Delta p}$. This leads to 
\begin{equation}
p_r=\Delta p = \sqrt{6 \left(U-\frac{1-M_0^2}{2}\right)},
\label{pr}
\end{equation}
which applies in the high-energy limit.  
The above prediction is compared with $N$-body simulations in Fig.~(\ref{fig7}): The velocity (and corresponding width) of the resonances is plotted 
for different energies $U$ (circles). The solid line refers to the analytical expression~(\ref{pr}). We notice a very good agreement between 
the numerics and the prediction~(\ref{pr}). As expected, as $U$ decreases, some discrepancy is observed since the system approaches the phase transition.

The above conclusion and in particular Eq.~(\ref{eq:HE}) can be also recovered using the following argument: In the high energy limit, the  
particles move essentially freely. The potential energy accommodates for just a small fraction of the total energy. Under this hypothesis, the individual phase 
$\theta$ evolves as: 
$$
\theta(t)=\theta_{0}+p_{0}t 
$$
where the index $0$ refers to the initial position of a single particle. From the definition of the magnetization, once a change of variables has been applied from $(\theta, p)$ to $(\theta_0,p_0)$, we obtain:
$$
M_n(t) \approx  \iint e^{in(\theta_{0}+p_{0}t)}f(\theta_{0},p_{0}) d\theta_{0} dp_{0},
$$
and Eq.~(\ref{eq:HE}) is recovered.

The next step is to incorporate the additional contributions, so far neglected. In particular, we focus our attention 
on the terms $t^{2n}\Delta p^{2n-2}$ in Tab.~\ref{tab:1}. For $\Delta\theta\approx \pi$ (i.e.\ $M_0\ll 1$), the dominant term is $-M_0\Delta p^2 t^4/36$ since 
all the other terms are of higher order in $M_0$. This latter can be seen as originated from a modification of Eq.~(\ref{eq:HE}) where a constant factor $c$ is 
being introduced as:  
\begin{equation}
 M(t)=M_{0}{\rm sinc}(t\sqrt{\Delta p^{2}-c}).
\label{sinc}
\end{equation}
The coefficient of $\Delta p^4 t^4$ of Eq.~(\ref{eq:HE}) is replaced by $\Delta p^4 t^4 -2c\Delta p^2 t^4$.  
Therefore $c=5/3$ matches the dominant term $-M_0\Delta p^2 t^4/36$, corresponding to the order $n=2$
.
The approximation of the magnetization given by Eq.~(\ref{sinc}) is in better agreement 
with the numerical simulations. In particular for the position of the resonances, Equation~(\ref{sinc}) gives $p_r=\sqrt{\Delta p^2 -c}$, which is closer to 
numerical values as shown Fig.~\ref{fig7} (dashed line). However, this additional term does not balance the analogous contributions associated with
higher orders ($n>2$) for which a slightly different value of $c$ is required. Deviations are however reasonably small (less than $10$ \%) over the range of inspected coefficients. The above argument can be extended to the case 
where $\Delta \theta < \pi$, so accounting for the terms 
proportional to $M_0$: In practice, an additional term of the type $c_1 M_0$ is introduced in the square roots of Eq.~(\ref{sinc}) where $c_1$ is a constant.  

\subsection{Low-energy limit}
\label{sec:LE}

We now consider the low-energy limit $U\simeq (1-M_0^2)/2$, that is $\Delta p\ll 1$. We notice that this limit is close to the line which marks the forbidden region in the parameter space $(M_0,U)$ (see e.g. \cite{chavanis08}). 
In what follows, we find an approximation of the coefficients of $M(t)$ proportional to $\Delta p^0$. 
We first observe that the Liouville operator~(\ref{eq:mh}) either increases or decreases by one order the exponent of $p$. Thus, the 
odd powers $\mh^k M$ contain a set of elements $b_{n,m}$ with $m$ odd. For the waterbag initial distribution, such terms vanish so $M(t)$ is an even function. 
Then, the recursion relation~(\ref{eq:taylor}) is generated by $\mh^2$. In the low-energy limit, if the kinetic terms are neglected, we get
\beq \mh^2 b_{u,0}\approx \frac{u}{2}\left( b_{1,0}b_{u-1,0}-b_{-1,0}b_{u+1,0} \right).\label{eq:h2LE} \eeq

\begin{figure}
    \center
    \includegraphics[width=8cm]{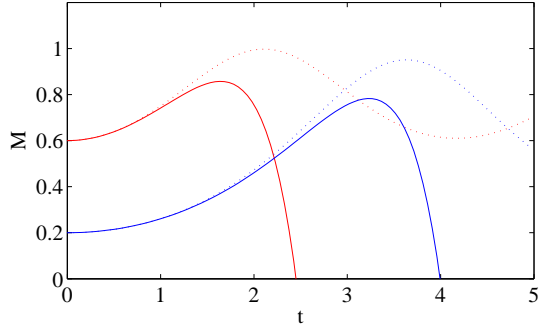}
    \caption{The magnetization $M(t)$ is plotted versus time, in the low-energy regime ($U=0.01$), for $M_0=0.2$ (blue curves) and $0.6$ (red curves). The solid lines refer to Eq.~(\ref{eq:MLE}) for $k_0=20$, the dotted ones to direct $N$-body simulations.}
    \label{fig:MLE}
 \end{figure}

An algebraic expression of the magnetization in the low-energy limit is obtained by studying the sequence of terms at the lowest order. In this way, we approximate $M$ as
\beq M(t)\approx \sum_{n=0}^{\infty} \left(\alpha_{n}b_{1,0}+\beta_{n}b_{-1,0}b_{2,0}+\gamma_{n}b_{-1,0}b_{1,0}^2\right)\frac{t^{2n}}{2n!}. \eeq
Using Eq.~(\ref{eq:h2LE}), we deduce that, at a given order $n+1$, the $b_{1,0}$ term comes from the $b_{1,0}$ term at order $n$, with $\alpha_{n+1}=\alpha_{n}/2$. 
From this recursion relation, we deduce the formula
\beq \alpha_{n}=\frac{1}{2^{n}}. \label{eq:alpha}\eeq
The $b_{-1,0}b_{2,0}$ term at the order $n+1$ is generated from both the $b_{1,0}$ and the $b_{-1,0}b_{2,0}$ terms at the lower order $n$. The recursion relation becomes $\beta_{n+1}=(\beta_{n}-\alpha_{n})/2$, which leads to
\beq \beta_{n}=-\frac{n}{2^n}. \label{eq:beta}\eeq
The third term in $b_{-1,0}b_{1,0}^2$ is not only generated through the reduced operator given by Eq.~(\ref{eq:h2LE}), but also from other nonlinearities: 
The latter terms have been neglected in Eq.~(\ref{eq:h2LE}), but appear when considering $\mh^4$ (and possibly higher powers of $\mh$) in the low-energy limit. 
We resort to an ansatz for $\gamma_{2n}$, fitting the coefficients derived algebraically up to $\gamma_{10}$:
\beq \gamma_{n}\approx -\frac{1}{9}\left(\frac{9}{2}\right)^n, \label{eq:gamma}\eeq
for large $n$.
It follows that the magnetization in the low-energy regime is approximated by
\begin{eqnarray} M(t)&\approx & M_0 \text{cosh}\left(\frac{t}{\sqrt{2}}\right)-M_0 M_2(0)\frac{t}{2\sqrt{2}}\text{sinh}\left(\frac{t}{\sqrt{2}}\right) \nonumber
\\ &&-M_0^3\frac{1}{9}\text{cosh}\left(\sqrt{\frac{9}{2}}t\right). \label{eq:MLE}\end{eqnarray}
This expression of the magnetization is compared with numerical simulations in Fig.~\ref{fig:MLE}. We notice the good agreement up to the saturation regime. 
As expected, for longer times, the approximation gets worse due to higher order nonlinearities which have been neglected. 

\subsection{Out-of-equilibrium phase transition}

As previously reported, increasing the energy $U$ at a fixed value of the initial magnetization $M_0$ leads to a drastic change in phase space 
which materializes as an out-of-equilibrium phase transition \cite{antoniazzi07} from an inhomogeneous to a homogeneous phase.
This phenomenon was first explained by invoking a principle of entropy maximization, based on the so-called theory of ``violent relaxation''~\cite{LB}. Another dynamical explanation of such transition comes from a bifurcation analysis in phase space~\cite{bach08}.

\begin{figure}
    \center
    \includegraphics[width=9cm]{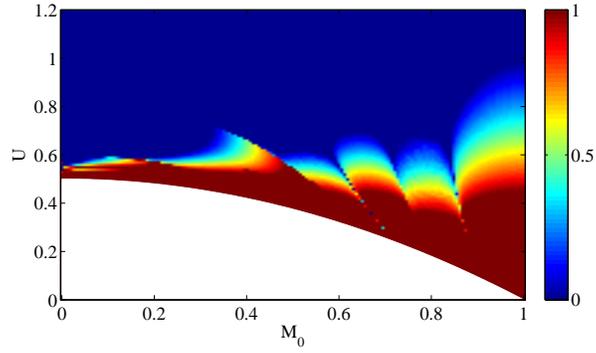}
        \caption{Map of the magnetization evaluated at the first local maximum of Eq.~(\ref{eq:taylor}) in the ($M_0,U$) plan. The data refer to the theoretical prediction
	calculated for $k_0=20$. The white region is the forbidden one.}
    \label{fig:phasetransition}
 \end{figure}

\begin{figure}
\center
\includegraphics[width=8cm]{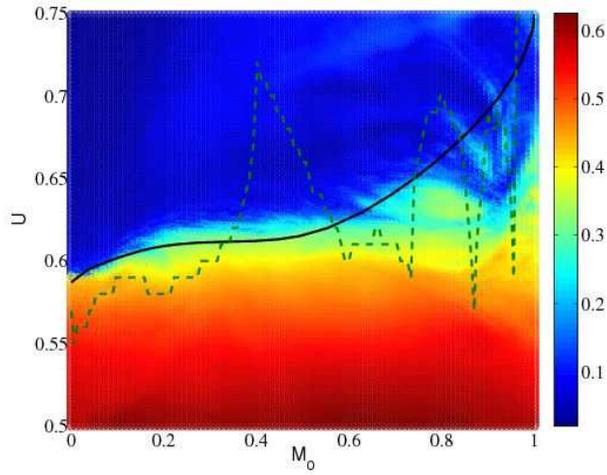} 
\caption{Map of the QSS magnetization in the ($M_0,U$) plan, as recorded via direct $N$-body simulations ($N=10000$). The solid line refers to the 
Lynden-Bell prediction. The dashed line stands for the bracket transition line with threshold magnetization set to $M_c=0.4$.}
\label{fig:phasetransition1}
\end{figure}

We here show that the transition can be also retrieved when tracking the short-time behavior of the magnetization. It means that the system relaxes very quickly in its metastable phase. The idea goes as follows: 
We monitor the magnetization dynamics via the analytical expression obtained from Eq.~(\ref{eq:taylor}) and store the first local maximum, for each choice of the pair ($M_0,U$). In case the series diverges, without passing through a local maximum, the intensity is recorded when its derivative crosses a given threshold (as a polynomial, it eventually explodes). We choose $k_0=20$ in the algebraic computations. The resulting values of the magnetizations are displayed in Fig.~\ref{fig:phasetransition} adopting a color code which continuously interpolates between the large ($M\approx 1$)
and small ($M\approx 0$) magnetization. As clearly shown, the upper portion of the parameter plane corresponds to almost homogeneous
configurations while magnetized phases are observed as the energy is reduced for fixed $M_0$. 
This scenario qualitatively agrees with direct numerical integrations, as confirmed by inspection of 
Fig.~\ref{fig:phasetransition1}. In the $N$-body simulations, the available parameter space ($M_0,U$) is partitioned in small cells, each associated with a reference water-bag distribution (that is, a two-level distribution). 
The QSS magnetization is measured by averaging the numerical time series over a finite time window after relaxation. The average QSS
magnetization is then represented using the same color code as above. When comparing 
Figs.~\ref{fig:phasetransition} and~\ref{fig:phasetransition1} it should be emphasized that the QSS regime occurs significantly after the violent relaxation process, beyond the first local maximum of the magnetization which is computed here. The results show that the average magnetization as recorded in the QSS correspond approximately to these local maxima in the non-homogeneous phase. A better quantitative matching can be obtained by considering higher order terms (larger $k_0$). Even though, improving the accuracy of the theoretical analysis is a crucial
requirement, already at this level of approximation it emerges a phase transition as clearly 
shown in Fig.~\ref{fig:phasetransition}.  

A reference line in parameter space ($M_0,U$) marking the transition between the magnetized and unmagnetized phases can be computed based on
the celebrated Lynden-Bell procedure, also known as the violent relaxation theory. The central idea of the Lynden-Bell approach consists in coarse-graining 
the microscopic one-particle distribution function $f(\theta,p;t)$ by introducing a local average in phase space. Starting from a waterbag initial profile 
with a uniform distribution $f_0$, a fermionic-like entropy can be rigorously associated with the coarse grained profile $\bar{f}$, namely 
$s[\bar{f}]=-\int \!\!{\mathrm d}p{\mathrm d}\theta \,\left[\frac{\bar{f}}{f_0} \ln \frac{\bar{f}}{f_0} +\left(1-\frac{\bar{f}}{f_0}\right)
\ln \left(1-\frac{\bar{f}}{f_0}\right)\right]$. The  corresponding statistical equilibrium, which applies to the relevant 
QSS regimes, is hence determined by maximizing such an entropy, while imposing the conservation of the Vlasov dynamical invariants: 
Energy, momentum and norm of the distribution. The analysis translates into the out-of-equilibrium phase 
transition line reported in Ref.~\cite{antoniazzi07}. Notice that the Lynden-Bell scenario recalled above formally applies to the waterbag initial
condition from which the fermionic principle is derived. Different energy functionals are at variance to be assumed when dealing with more 
complex initial conditions and there is no a priori guarantee that the maximum entropy strategy would perform equally well. Aiming at extracting a transition line from the 
viewpoint of the bracket calculation, one could impose a critical threshold $M_c$ to the magnetization: First local maximum values of the magnetization larger than $M_c$ are assumed 
to yield a magnetized QSS, while for magnetization below the reference value $M_c$ the system evolves toward a homogeneous QSS. The (arbitrary) choice $M_c=0.4$ leads to a transition line (dashed line in Fig.~\ref{fig:phasetransition1}) 
which resembles qualitatively the Lynden-Bell line (solid line). Notice that magnetized patches are numerically seen to extend over the region of homogeneous QSS, so
effectively deforming the transition boundary in a non trivial way. Interestingly, such islands are entrapped in the wiggles of the bracket 
transition profile. 

In conclusion, the bracket method returns sensible information on the existence of an out-of-equilibrium transition, so 
resulting in a powerful tool for those generalized settings where the Lynden-Bell ansatz proves inadequate (as for instance, for Gaussian initial 
conditions) or, at least, cumbersome (e.g.\ multi-level initial distribution). 

\section{Conclusion}

In this paper we have focused on a paradigmatic Hamiltonian mean-field model, often being investigated for its long-living Quasi Stationary States (QSS), and for its peculiar out-of-equilibrium dynamics and phase transitions. Starting from an out-of-equilibrium initial 
conditions of the waterbag type, the system rapidly evolves toward an intermediate dynamical regime, distinct from the corresponding equilibrium configuration.
When increasing the number of interacting elements the time to equilibration gets longer and formally diverges when the thermodynamic limit is performed. Aiming at
shedding light onto the short time dynamics, which ultimately governs the QSS emergence, we have here resorted to an analytical approach. The idea is to 
develop an algebraic technique based on the Lie-Poisson structure of the HMF dynamics. In doing so we are able to return an analytical
prediction for the global magnetization as a function of time, a macroscopic parameter sensitive to the microscopic particles
evolution. Two limiting cases are explicitly considered, respectively the high and low energy settings, and shown to yield to tractable
expressions for the magnetization amount. In general, and due to the perturbative nature of the calculation, the full analytic expression contains a vast
collection of terms which are difficult to handle. The number of terms involved increases rapidly with the order of the approximation making it practically difficult
to address the dynamics in the relevant, saturated, QSS regime.  However, targeting the analysis to the first 
local maxima in the magnetization, and accounting for $20$ orders in the perturbative expansion, the existence of an out-of-equilibrium phase transition was singled out, 
separating between homogeneous and non-homogeneous QSS. This transition was already recognized in  Ref.~\cite{antoniazzi07} and interpreted using an {\em ad hoc} maximum entropy principle 
suited for waterbag initial profiles. Although the calculations are carried out for the so-called waterbag initial condition, the technique we use in this article is rather flexible and can be readily extended to 
other, possibly more general classes of initial conditions so returning fully predictive scenarios. We also notice that the proposed method can be adapted to other contexts where long-range many body interactions are at play. The method is particularly adapted to short-time dynamics (transients, metastable states, violent relaxation, etc...).

\section*{Acknowledgments}
CC acknowledges useful discussions with the Nonlinear Dynamics team of the CPT. This work is supported by Euratom/CEA (contract EUR 344-88-1 FUA F). 
Y.Y.Y. has been supported by the Ministry of Educations, Science, Sports and Culture, Grant-in-Aid for Young 
Scientists (B), 197660052.

\end{document}